\documentclass[%
 reprint,
 amsmath,amssymb,
 aps,
showkeys,
longbibliography,
]{revtex4-2}

\usepackage{graphicx}   
\usepackage{dcolumn}    
\usepackage{bm}         
\usepackage{float}      
\usepackage{xcolor}
\usepackage{amssymb}
\usepackage[ruled, lined]{algorithm2e}

\SetCommentSty{mycommfont}
\usepackage[english]{babel}

\usepackage[mathlines]{lineno}

\begin{document}

\preprint{APS/123-QED}

\title{An Efficient MCMC Approach to Energy Function Optimization \\in Protein Structure Prediction}

\author{Lakshmi A. Ghantasala}
\email{lghantas@purdue.edu, Corresponding Author}
\affiliation{Elmore Family School of Electrical and Computer Engineering, Purdue University}
\author{Risi Jaiswal}
\affiliation{Elmore Family School of Electrical and Computer Engineering, Purdue University}
\author{Supriyo Datta}
\affiliation{Elmore Family School of Electrical and Computer Engineering, Purdue University}


\begin{abstract}
 Protein structure prediction is a critical problem linked to drug design, mutation detection, and protein synthesis, among other applications. To this end, evolutionary data has been used to build contact maps which are traditionally minimized as energy functions via gradient descent based schemes like the L-BFGS algorithm. In this paper we present what we call the Alternating Metropolis-Hastings (AMH) algorithm, which (a)	significantly improves the performance of traditional MCMC methods, (b) is inherently parallelizable allowing significant hardware acceleration using GPU, and (c)	can be integrated with the L-BFGS algorithm to improve its performance. The algorithm shows an improvement in the energy of 8.17\% to 61.04\% (average 38.9\%) over traditional MH and 0.53\% to 17.75\% (average 8.9\%) over traditional MH with intermittent noisy restarts, tested across 9 proteins from recent CASP competitions. We go on to map the Alternating MH algorithm to a GPGPU which improves sampling rate by 277x and improves simulation time to a low energy protein prediction by 7.5x to 26.5x over CPU. We show that our approach can be incorporated into state-of-the-art protein prediction pipelines by applying it to trRosetta2’s energy function and also the distogram component of Alphafold1’s energy function. Finally, we note that specially designed probabilistic computers (or p-computers) can provide even better performance than GPU's for MCMC algorithms like the one discussed here.

\end{abstract}

\maketitle


\section{\label{sec:1}Introduction}

The problem of protein folding is a decades long quest launched in large part by Anfinsen’s thermodynamic hypothesis that a protein’s geometry can be inferred from only its’ amino acid sequence \cite{anfinsen_principles_1973}. In the last decade, co-evolutionary sequence data for a given protein has been used to build contact maps from which the 3-D geometry of a protein can be inferred. Contact maps constitute a set of probabilities for inter-residue distances or orientations.  The structures that satisfy these probabilities most are then the predictions for the 3-D geometry of the protein for which the contact map is intended \cite{senior_improved_2020,yang_improved_2020}. While evolutionary couplings may be statistically derived \cite{marks_protein_2011, hopf_sequence_2014}, deep neural networks have seen resounding success in predicting a protein’s contact map \cite{di_lena_deep_2012,li_respre_2019,zheng_detecting_2019}. Deep neural networks were notably used by the Alphafold1 team to come in first in CASP13 (2018) \cite{alquraishi_alphafold_2019}, and again by trRosetta2 in CASP14 (2020). 

Predicting highly accurate contact maps from co-evolutionary data, i.e optimizing inter-residue distance and orientation probabilities, is the first half of the problem. These probabilities may then be formulated into an energy function whose minimum describes the optimal conformation of the protein. The goal is to find a set of residue ($C_\beta$) positions that minimize the derived energy function and related force-fields. For both Alphafold1 and trRosetta2, non-monotone L-BFGS \cite{liu_limited_1989}, a second order gradient descent method, was used. On the other hand, it has recently been shown that sampling approaches may be better suited to find optimal states in non-convex energy landscapes when compared to traditional gradient descent approaches \cite{ma_sampling_2019}. This naturally raises the suspicion that a direct sampling approach to minimizing contact map-based energy functions may lead to better results than gradient descent. 

We present a novel sampling approach termed Alternating Metropolis-Hastings (Alternating MH) that acts directly on the torsion angles (ex.~$\varphi$, $\psi$) of a protein to find optimal conformations. Alternating MH improves upon traditional MH by introducing local energies by which proposals can be judged. Localized MH limits the scope of the energy to short segments within the full protein. By limiting the scope of the energy function, we essentially assign a ‘local’ energy to a proposal state that is based only on the immediate surroundings of the modified variable. The current energy is also determined on the same range, and the comparison now judges whether the proposal improves local energy or not. Alternating MH 'alternates' between local and global MH to optimize an energy function. 

It should be noted that recent works like Alphafold 2 \cite{jumper_highly_2021} and RoseTTAFold \cite{baek_accurate_2021} based on end-to-end prediction of protein structure using transformer networks have observed remarkable success. These machine learning pipelines have made significant improvements over the previous contact-map based methods that our Alternating MH is shown to expedite. However, our method should be adaptable to new state-of-the-art techniques. Indeed it may be even more useful in future for $\it{in \ vivo}$ protein structure prediction which will require sampling from stochastic environments as opposed to the controlled environments encountered $\it{in \ vitro}$.

\section{\label{sec:2}Methods}

\subsection{\label{sec:2.1}Alternating Metropolis-Hastings Algorithm}

\begin{figure*}[t]
\includegraphics[width=\linewidth]{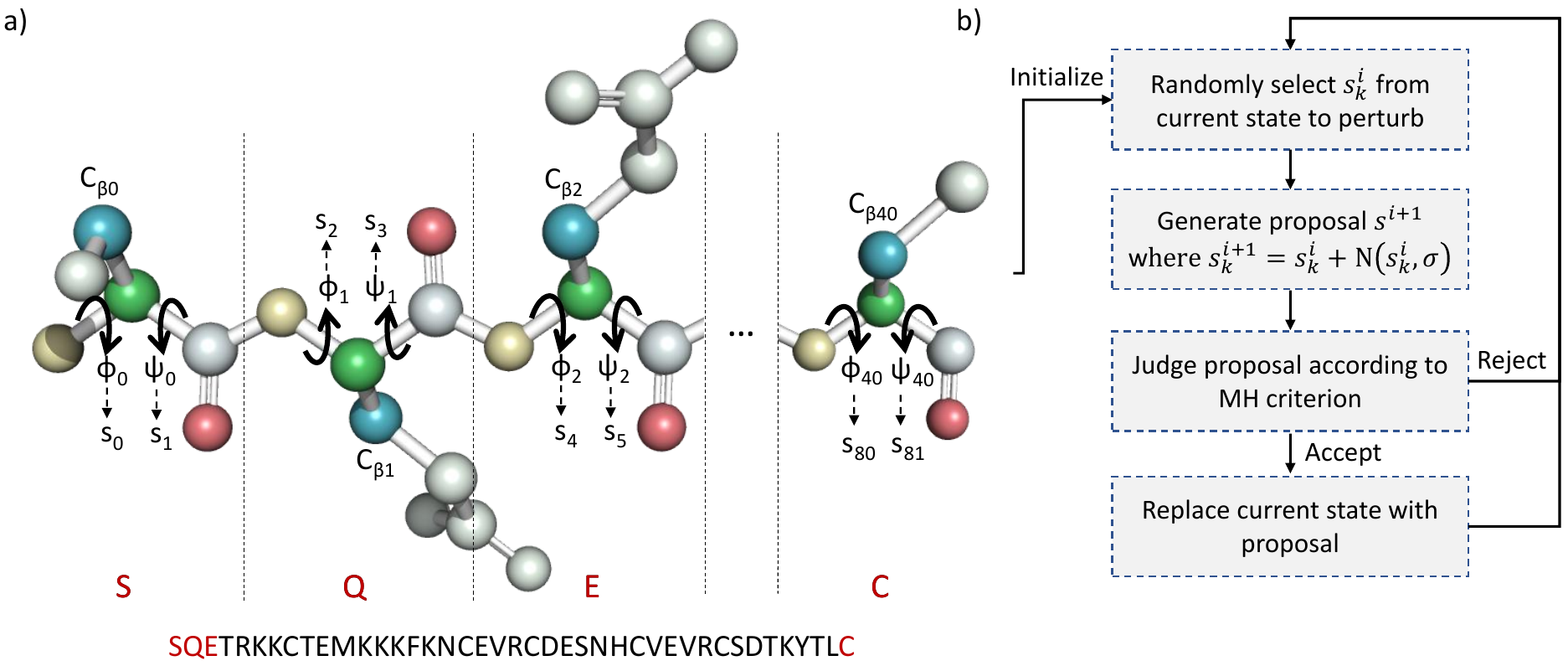}
\caption{\label{fig:map_protein_to_mh} (A) The mapping of an unfolded protein (T0955 from CASP13) to variables is shown here. The dihedral angles $(\varphi, \psi)$ are captured by a single vector s such that $s^i_{\textrm{even}} = \varphi^i_{1,\ldots,r}$ and $s^i_{\textrm{odd}} = \psi^i_{1,\ldots,r}$. The uncolored atoms attached to the $C_\beta$ (blue) atoms are the side chains for each residue. The unfolded protein in (A) is the initial ‘current state’ in the following flow chart. (B) depicts the flow chart for an MH based minimization process. Alternating MH impacts step 3, where it modifies the energy used for the MH criterion between traditional MH and localized MH.}
\end{figure*}

Traditional Metropolis-Hastings (MH) forms the first stage of Alternating MH. Proposal states are repeatedly generated, beginning at an input initial state, and sequentially accepted or rejected according to an acceptance probability ($P_{accept}$). The proposal scheme acts on an $n$-variable state at step $i$, $s^i_{1,\ldots,n}$ , where $s^i_{even}=\varphi^i_{1,\ldots,r}$ and $s^i_{odd}=\psi^i_{1,\ldots,r}$ for a protein with $r$ residues. This mapping, used by Alphafold1, is illustrated in Fig.~\ref{fig:map_protein_to_mh}. While different implementations may use different mappings (ex. trRosetta2 follows a 5 angle parameterization), they may all be mapped to the $s$ vector, which is finally used to generate proposals. The proposal mechanism is a very simple 'single-flip' mechanism where a Gaussian noise is added to a single torsion angle in $s$ per proposal. After $n$ proposals, we arrive at an entirely new sample. Energy is calculated as a function of $s$ and the coordinates of atoms in the resulting structures.  The probability of acceptance of a proposal for traditional MH is described in Eq.~\ref{eq:mh}:

\begin{subequations}
\label{eq:all_mh}
\begin{eqnarray}
P_{accept}&&(s^i_{1,\ldots,n}) = \nonumber\\&&\min \Big[1, \frac{P(s^{i+1}_{1,\ldots,n}) P(s^{i+1}_{1,\ldots,n}|s^{i}_{1,\ldots,n})}{P(s^{i}_{1,\ldots,n}) P(s^{i}_{1,\ldots,n}|s^{i+1}_{1,\ldots,n})} \Big] 
\label{eq:mh}
\end{eqnarray}

where superscripts define the iteration of the state and subscripts define the range of variables being considered within the state. The subscripts will become relevant when a localized version of this algorithm is defined. We can simplify the acceptance criterion to remove dependence on the conditional probability, as a symmetric, Gaussian transition distribution is used: 

\begin{eqnarray}
P(s^{i+1}_{1,\ldots,n}|s^{i}_{1,\ldots,n}) = N(s^i_k,\sigma)
\label{eq:mh_transitional}.
\end{eqnarray}

The marginal component of the acceptance criterion may be defined as:

\begin{eqnarray}
P(s^{i}_{1,\ldots,n}) = e^{-V(x)}
\label{eq:mh_marginal}.
\end{eqnarray}

where $x = G(s^i_{1,\ldots,n})$ defines the coordinates ($x$) for all $C_\beta$ atoms in the protein as a function of the torsion angles of the chain.
\end{subequations}

Alternating MH is a combination of two stages. Stage one is traditional MH as described in Eqs.~\ref{eq:all_mh}. Stage two is termed Localized Metropolis-Hastings (see Eqs.~\ref{eq:all_local}), which uses a modified energy function that assigns energy to a proposal based only on some local segment of $s$. The acceptance of a proposal is now based on the energy of the segment of the residue chain:
\begin{subequations}
\label{eq:all_local}
\begin{eqnarray}
P_{accept}(s^i_{1,\ldots,n}) = \min \Big[1, \frac{P(s^{i+1}_{k-d,\ldots,k+d})}{P(s^{i}_{k-d,\ldots,k+d})} \Big] 
\label{eq:localizedmh}
\end{eqnarray}

where $k$ is the index of the variable being modified within the state and $d$ is the radius of the segment. To get the marginal probability of a segment of a proposal, a new localized potential must be defined which returns the segment's energy:

\begin{eqnarray}
P(s^{i}_{k-d,\ldots,k+d}) = e^{-V_{local}(x, k, d)}
\label{eq:mh_marginal}
\end{eqnarray}

Generally, $V$ requires summing over all inter-residue (and inter-atomic) energies for a protein; we may define a localized energy function $V_\textrm{local}$ by summing over a subset of the inter-residue energies defined by $k$ and $d$.  
\end{subequations}

\begin{figure}[H]
\includegraphics[width=0.5\textwidth]{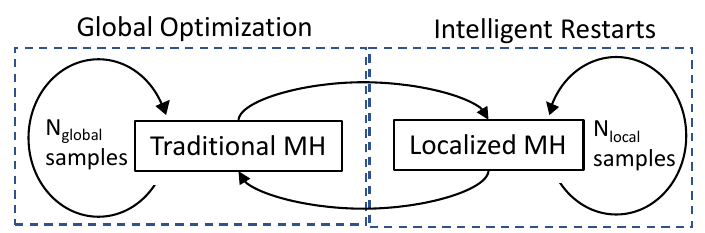}
\caption{\label{fig:minisummary} The Alternating Mh scheme cycles between generating $N_{global}$ samples via traditional MH, and $N_{local}$ samples via a novel localized MH. The traditional MH may be thought of as the global optimization component of the process, akin to L-BFGS, while the localized MH may be compared to 'noisy restarts', an intelligent way to escape from local minima.} 
\end{figure}

Localized MH follows the same Gaussian transition probability defined in Eq.~\ref{eq:mh_transitional}. Alternating MH alternates between local and global MH, judging $N_{local}$ proposals according to local energies, then $N_{global}$ proposals according to global energy, as in Fig.~\ref{fig:minisummary}. The pseudo code for the Alternating MH scheme is shown in Alg.~\ref{alg:altmh}. 


    

\begin{algorithm}
\caption{Alternating Metropolis-Hastings} \label{alg:altmh}
         \SetKwInOut{Input}{input}
        \SetKwInOut{Output}{output}
        \Input{$i = 0, d = 10, s^0=\{180^{\circ},\ldots, 180^{\circ}\}$}
        \Output{$s \gets \textrm{optimal angles}$}
        
\For{$l \gets 0$ to $cycles$}{
    \tcc{Traditional Metropolis-Hastings}
    \For{$i \gets 1$ to $N_{\textrm{global}}$}{ 
        \For{$k \gets 1$ to \textrm{length}($s$)}{
            $s^{'} \gets s^{i-1}$\\
            $s^{'}_{k} \gets s^{i-1}_{k} + N(0,\sigma)$\\
            $P_\textrm{accept} = \min \Big[1, \frac{e^{-V(\textrm{G}(s^{i-1}))}} {e^{-V(\textrm{G}(s^{'}))}} \Big]$\\
            \uIf{$\textrm{rand}(0,1) < P_\textrm{accept}$}{
                $s^i \gets s^{'}$
            }
            \Else{    
                $s^i \gets s^{i-1}$
            }
        }
    }
    \tcc{Localized Metropolis-Hastings}
    \For{$i \gets 1$ to $N_{\textrm{local}}$}{
        \For{$k \gets 1$ to \textrm{length}($s$)}{
            $s^{'} \gets s^{i-1}$\\
            $s^{'}_{k} \gets s^{i-1}_{k} + N(0,\sigma)$\\
            $P_\textrm{accept} = \min \Big[1, \frac{e^{-V_{\textrm{local}}(\textrm{G}(s^{i-1}), k, d)}} {e^{-V_{\textrm{local}}(\textrm{G}(s^{'}), k, d)}} \Big]$\\
            \uIf{$\textrm{rand}(0,1) < P_\textrm{accept}$}{
                $s^i \gets s^{'}$
            }
            \Else{
                $s^i \gets s^{i-1}$
            }
        }
    }
}
\end{algorithm}

\subsubsection{Localization Radius Parameter $d$}

The localization radius parameter describes the number of variables in $s$ from variable $k$ that we may consider for the MH criterion. This gives us a powerful knob with which to tune how the MCMC procedure folds the protein. The smaller $d$ is, the fewer residues each proposed fold has to satisfy, and the larger the acceptance rate is. Traditional MH, by this metric, has the smallest acceptance ratio where every proposal must satisfy, to some degree, every residue in the chain. 

As the protein folding problem may be captured by a 1-D state array ($s$), $d$ amounts to looking to the left and right of the $k^{th}$ variable in $s$. If the state vector is 2 (or more)-Dimensional, the segment may grow as a circle ($n$-d sphere), radiating outwards from the center $k$ variable with radius $d$. The advantage then of defining a radius parameter, rather than a range parameter, is in easily extrapolating this algorithm to problems with higher dimensional state vectors. 

One can cascade multiple localization ranges to consider folds of multiple sizes. While results for this work alternate between global MH, i.e $d=2n$ where $n$ is the length of $s$, and a localized MH with $d=10$, further ranges could also be included. Intuitively, it would seem the localization ranges should reflect the sizes of local structures in the problem, though future work will have to determine how best to cater the radius to the problem at hand. 

\subsection{\label{sec:2.3} Applying Alternating MH to Protein Structure Prediction}

Two implementations of the Alternating MH scheme are demonstrated for contact map-based energy minimization. The first implementation (described by Eq.~\ref{all_af}) minimizes an energy function built with Alphafold1’s published distograms \cite{senior_improved_2020, senior_protein_2019}. The energy function is as follows: 

\begin{subequations}
\label{all_af}

\begin{eqnarray}
V(x) = -\sum_{i,j,i\neq j} \log P(|x_i - x_j| | S, MSA(S))
\label{eq:af_full}
\end{eqnarray}

Where $x=G(\varphi, \psi)$, $S$ is the sequence of amino acids to be folded, and $MSA(S)$ is the Multiple Sequence Alignment of that sequence. This function is a part of the total energy used in the Alphafold1 pipeline, which also includes a torsion potential and a physics based term that includes van-der-waals forces; these additional terms are discounted in this paper to make an accelerated GPGPU implementation more reliable. To find the energy for a localized segment, the log probabilities are summed over only within the desired range:

\begin{eqnarray}
\label{vlocal_alphafold}
&&V_{\textrm{local}}(x, k, d) =\nonumber \\ && -\sum^{\lfloor \frac{k}{2} \rfloor + \lfloor \frac{d}{2} \rfloor}_{i,j= \lfloor \frac{k}{2} \rfloor - \lfloor \frac{d}{2} \rfloor, i\neq j} \log P(|x_i - x_j| | S, MSA(S))
\label{eq:af_local}
\end{eqnarray}

where $k$ is the index of the variable modified between subsequent states, and $d$ is the radius of the segment within the angle vector ($s$). As there are 2 torsion angles per residue in the parameterization described by Alphafold1, the angle being modified is mapped to its corresponding $C_\beta$ atom in the chain by dividing by 2 and flooring. 

Details regarding the GPGPU implementation of the minimization of this distogram-only energy function are in section~\ref{sec:2.4}.

\end{subequations}

The second implementation, described in Eq.~\ref{eq:tr_full} minimizes an energy function that is a part of the trRosetta2 pipeline \cite{yang_improved_2020}. Unlike with Alphafold1, the complete energy function was minimized with Alternating MH, which includes both restraint and physical forces:

\begin{eqnarray}
V(s) =&& V_{\textrm{distance}}(G(s)) + V_{\textrm{torsion}}(s) + \\ &&V_{\textrm{rosetta}}(G(s)) \nonumber
\label{eq:tr_full}
\end{eqnarray}

where  $V_\textrm{rosetta}$ includes energies corresponding to ramachandran (rama), hydrogen bonding (cen\_hb), and van der Waals (vdw) forces \cite{yang_improved_2020}. A localized version of this function was derived via the $get\_sub\_score$ function in pyrosetta. The trRosetta2 pipeline parameterizes $C_\beta$ atoms according to $(\varphi, \theta, \omega)$, which is  different than Alphafold’s parameterization, though once the information is captured in the angle vector ($s$), the remaining procedure is unchanged. Proposals are still generated and judged via $s$.

The original minimization process for trRosetta2 involves modifying the weights of the score function as the minimization scheme iterates; for the sake of an exact comparison between L-BFGS and Alternating MH, the process was simplified to minimizing a single, constantly weighted energy function. Furthermore, the final relaxation procedure (FastRelax) is not considered. The final RMSD values, then, will not match with trRosetta2's published results, but should make for a robust comparison with the Alternating MH scheme.

In both implementations, $s$ is initialized to \{180\textdegree\}, as shown in Fig.~\ref{fig:map_protein_to_mh}. The proposal scheme picks $s^i_k$ to perturb by a 0 mean Gaussian noise at each proposal, incrementing $k$ from 0 to $n$ where $n$ is the length of $s$. The variance of the Gaussian ($\sigma$) is gradually reduced over the course of an Alternating MH run; for this work, $\sigma = 10 \rightarrow 4$. This ensures that initial proposals result from larger swings in $s$, and later proposals become tighter. The full Alt-MH procedure was repeated 10 times per tested protein, where each restart would begin at the lowest energy found during the previous run. If a lower energy was not found during the previous Alt-MH run, the range of $\sigma$ changes to $10/L \rightarrow 1/L$ where $L$ is the number of runs where a lower energy was not found.

\begin{figure*}
\includegraphics{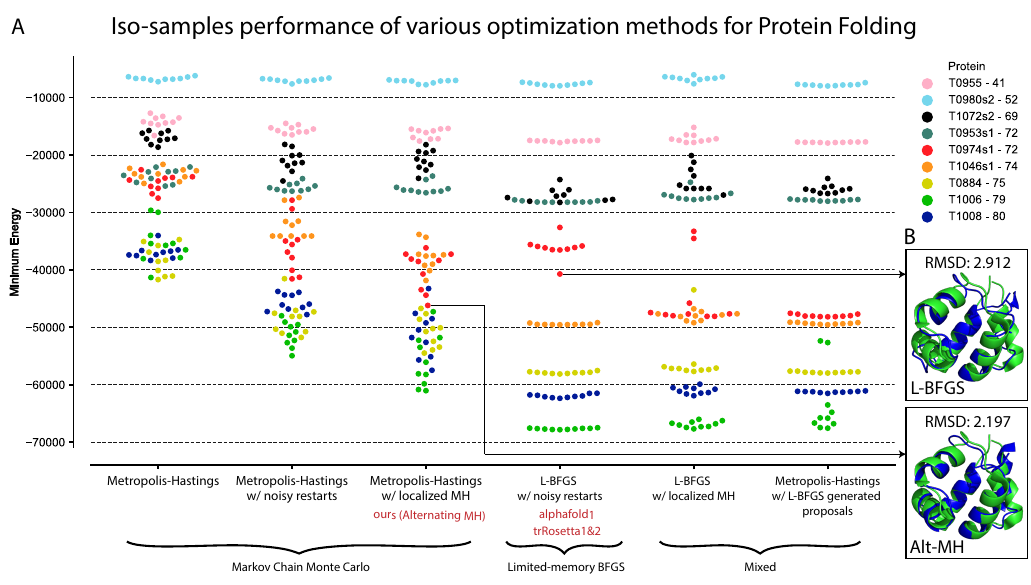}
\caption{\label{fig:isosamples} (A) The Alternating Metropolis-Hastings method presented in this paper (column 3) is benchmarked against various optimization schemes that include MCMC, L-BFGS, and mixed schemes. Each scheme is applied 10 times to each of 9 different proteins sampled from the previous 3 CASP competitions; the minimum energies observed over each run are plotted. All runs are for 10000 samples. While the L-BFGS implementation in column 4 is based specifically on trRosetta2's implementation, Alphafold1 and trRosetta1 also use a variation of this approach, hence their inclusion in the label. (B) The lowest energy prediction for T0974s1 by the standard L-BFGS scheme (blue) and our proposed Alternating MH scheme (blue) shows Alt-MH is able to find a lower energy state with improved RMSD to the original structure (green).}
\end{figure*}

\subsection{\label{sec:2.4} Hardware Acceleration of Alternating MH}
The Metropolis Hastings algorithm is generally amenable to hardware that allows multiple \textit{chains} to be run in parallel. The more chains the hardware accommodates, the larger the space that is searched and consequently the better the chances of finding lower energy ground states. 

A General Purpose Graphics Processing Unit (GPGPU) was chosen as a suitable hardware accelerator for Alt-MH, as it is a parallel computing platform that naturally accommodates the complex floating-point operations required for this problem (ex. cos, sin, log, sqrt), as opposed to an FPGA. As a note, we use the terms GPGPU and GPU interchangeably in this manuscript as the Nvidia P100 may be seen as both. A simplified diagram of the GPU architecture as an accelerator for the Alternating MH algorithm is presented in \ref{fig:gpu}. The logical blocks in the architecture of a p-computer~\cite{kaiser_benchmarking_2021}, shown in \ref{fig:gpu}.b, can be matched 1-to-1 with our mapping of AMH to a GPU. 

Each block in the GPU houses an MCMC chain while an array of threads within the block parallelize the operations \textit{within} an MCMC chain. These two forms of parallelism are key to achieving the presented performance. 
\subsubsection{Intra-Block parallelism}
Alt-MH, and in-fact MCMC algorithms in general, involve(s) 4 steps; random number generation, proposal generation, energy calculation, and criterion evaluation; these steps may be parallelized to observe faster execution. For the protein structure prediction problem, there are 3 steps that may take advantage of intra-block parallelism:

\textbf{1. RNG}: Random numbers are required for a) generating a proposal, and b) evaluating the MH criterion. To generate these, each MCMC block in the GPU maintains two LFSRs, assigned to the first 2 threads in each block. Each LFSR produces a pseudo-random, uniformly distributed random number. One of these is converted to a Gaussian-distributed random number via the Box-Muller method, as required for proposal generation. 

\textbf{2.	Proposal Generation}: A proposal protein structure is generated by adding a Gaussian distributed, 0 mean value to each $\varphi$ or $\psi$ bond angle in the protein. Every time a bond angle is modified, the subsequent (if the bond-of-interest is phi) or preceding (if the bond-of-interest is psi) atoms must rotate around the axis of the bond-of-interest. On CPU, this rotation is done atom by atom; on GPU, each atom is assigned to its own thread, which allows rotation of atoms in parallel.  Atoms are rotated $n$ at a time, where $n$ is the length of the angle vector $s$. 

\textbf{3.	Energy Calculation}: The Alphafold1 implementation of Alt-MH relies on the calculation of pairwise energies between residues; pairwise energies are then summed across all ij residue pairs that satisfy $\lfloor \frac{k}{2} \rfloor - \lfloor \frac{d}{2} \rfloor < i < j < \lfloor \frac{k}{2} \rfloor + \lfloor \frac{d}{2} \rfloor$, where k is the residue housing the bond-of-interest and $d$ is the localization range. In the GPU, each thread determines the pairwise energy of a single ij pair. Each thread maintains a sum of its' ij pairs until all pairwise energies are accounted for. Then parallel reduction is employed to efficiently sum the resultant energy terms across every thread in the block.

\begin{figure*}
\includegraphics[width=\linewidth]{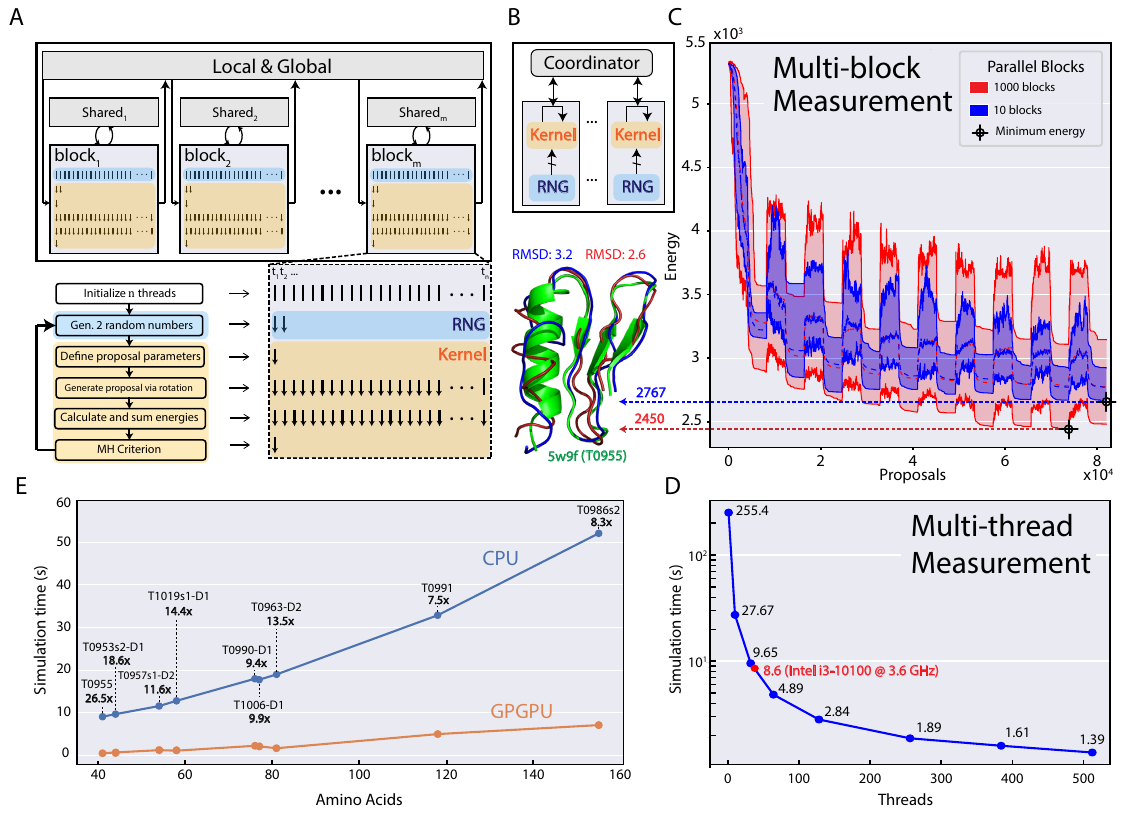}
\caption{\label{fig:gpu} (A) describes how the Alternating-MH algorithm may take advantage of parallelized hardware to improve performance. The RNG and Kernel described by the generic p-computer architecture in (B) can be found inside each MCMC block. The operations within an MCMC block may be parallelized via threads in the GPU as shown in (A). (C) measures the improvement in minimum energy found via 1000 parallel blocks over 10 parallel blocks. While the mean (dotted lines) of the two runs largely overlap, the min and max (solid lines) widen between the 10 and 1000 block runs. (D) measures the effect of the number of threads in a block on simulation time. (E) notes an overall improvement in simulation time between the CPU (Intel I3-10100 @ 3.6Ghz) and GPU (Nvidia P100) implementations across 9 protein domains, annotated with their CASP labels.} 
\end{figure*}

The effectiveness of parallelizing intra-block operations is shown in Fig~\ref{fig:gpu}.e. As the number of threads grows, the simulation time of reaching 8.2e4 proposals reduces.  

\subsubsection{Inter-Block parallelism}
Multiple Alt-MH blocks each with their own RNG will explore different slices of the state space. The wider the search, the better the chances of finding lower energy states. Accomodation of multiple parallel blocks also opens the door for various coordinated MCMC schemes to enter the picture, ex. parallel tempering~\cite{hansmann_parallel_1997}, multiple-try metropolis~\cite{liu_multiple-try_2000}, evolutionary monte-carlo~\cite{liang_evolutionary_2000}, and interchain adaptive MCMC~\cite{craiu_learn_2009} to name a few. These methods may work hand in hand with Alt-MH. Alt-MH deals with improving a single MCMC block while these methods work to coordinate multiple MCMC blocks to gain some advantage.

\section{\label{sec:3}Results}

The trRosetta2 based implementation is used as a benchmark to compare Alt-MH against standard MCMC approaches, nonmonotone L-BFGS, and mixed approaches, as seen in Fig.~\ref{fig:isosamples}. Each optimization scheme was run for 10,000 samples. A sample in the context of L-BFGS is when every variable in the state has been perturbed according to its' gradient once. A sample in the context of MH is when every variable in the state has had a chance to be perturbed once. Note that for MH based schemes, a sample need only be a full set of \textit{attempted} perturbations, as opposed to gradient descent based schemes where a sample is a full set of perturbations.

The first column shows traditional Metropolis-Hastings with annealing, i.e $0.1 < \beta < 1$ where $\beta$ is inverse temperature. The second column is traditional MH, with the same annealing, but with a noisy restart every 100 samples. The noisy restarts add a uniform random number between -10 and 10 to every torsion angle in the protein. Column 3 is the Alternating-MH algorithm presented in this paper. The benchmark was done with 80 samples of global-MH and 20 samples of localized MH with localization radius $d=5$. Perturbation variance for proposals was reduced from $10^{\circ}$ to $4^{\circ}$ over the course of each run, so later proposals are made within a shorter distance from the current state than earlier proposals. This method improves upon the energies of Traditional MH (column 1) by 8.17\% for T0980s2 to 62.27\% for T1046s1 (and 61.04\% for T1006) with average improvement at 38.9\%. We also observe improvement over MH with noisy restarts (column 2) of 0.53\% for T0953s1 to 17.75\% for T1046s1.  

The L-BFGS method in the benchmark is the lbfgs\_armijo\_nonmonotone minmover from pyrosetta, with implementation details matching trRosetta2's code. Because this is a nonmonotone optimization scheme, it was benchmarked by running in batches of 100 rather than in individual steps; performance drops significantly when run one iteration at a time. Column 5, L-BFGS w/ localized MH, uses the same L-BFGS function as column 4, but shortens the L-BFGS samples to 80 and replaces the noisy restarts with 20 samples of localized MH, the scheme described in this manuscript. Despite running for 20\% fewer samples of L-BFGS, column 5 exhibits predictions largely on par with the pure L-BFGS method, i.e to within 3.5\% for 7 of the 9 proteins, and within 12.8\% worst case. The last method, Metropolis-Hastings with L-BFGS generated proposals, implements a Metropolis-adjusted L-BFGS algorithm in a similar fashion to the well known Metropolis-adjusted Langevin algorithm (MALA)~\cite{grenander_representations_1994}. Proposals are generated via lbfgs\_armijo\_nonmonotone run for 20 samples, and accepted or rejected according to the metropolis-hastings criterion 10000 times. This approach proved to be exactly on par with the standard L-BFGS approach (column 4), coming to within 1\% worse than L-BFGS for 5 proteins (worst case 6.4\% for T1006), and beating L-BFGS by 1\% for 2 proteins.

There is one odd-ball result that stands out amongst all the tested proteins, T0974s1. Every scheme with some MH component finds a lower energy state for T0974s1 compared to the standard L-BFGS minimization. Future work would have to discover the quality that causes this. For T0974s1, our algorithm outperforms the standard L-BFGS approach by 10.51\% in energy and 32.5\% in RMSD, as seen in Fig.~\ref{fig:isosamples}.B. L-BFGS with Localized MH (Column 4) outperforms the standard approach by 23.73\% in energy, and Metropolis-adjusted L-BFGS (Column 5) outperforms by 32.34\%.

The Alternating-MH implementation described in Eq.~\ref{all_af} was mapped to an NVIDIA P100 GPU running on Google Colaboratory, the results of which are shown in Fig.~\ref{fig:gpu}. Fig.\ref{fig:gpu}.A describes how the algorithm fits into the architecture of a GPU to take advantage of the hardware's parallelism. The architecture of a p-computer can be overlaid on the GPU architecture to show that this GPU implementation fits into the Kernel-RNG paradigm for a p-computer described by Kaiser et al~\cite{kaiser_benchmarking_2021, kaiser_probabilistic_2021}, shown in \ref{fig:gpu}.B. 

Fig~\ref{fig:gpu}.C showcases the benefits of having multiple parallel MCMC blocks; The 1000 parallel block run finds a structure with 12.97\% lower energy than 10 blocks run in parallel; this directly translates into a predicted protein structure with 23.0\% lower RMSD to the original protein. Future works may explore coordinated MCMC schemes that may lead to significant improvements in larger, multi-block runs. The GPU also allows for intra-block parallelism, dependant on the number of threads assigned to a block. Fig.~\ref{fig:gpu}.D shows how the number of threads correlates with the simulation time of 82000 proposals of Alt-MH for T0955. Finally, both forms of parallelism lead to lower overall simulation time for the GPU implementation compared to the CPU implementation, shown in Fig.~\ref{fig:gpu}.E. The CPU implementation (Intel I3-10100) was run for 1e5 proposals. The average time over 10 runs is plotted and the average minimum energy was noted. The GPU was then run until a structure with the same minimum energy is found for each protein. 

This approach is intended to be similar to the time to solution metric described in other literature \cite{patel_ising_2020, kaiser_benchmarking_2021, kaiser_probabilistic_2021}; here, there is an additional challenge that the true ground state cannot be reached in reasonable time (or likely ever due to error in the energy function). Hence we match GPU to CPU optimality, rather than matching both to an optimality gap. Overall we observe improvements in time to solution of 7.5x to 26.5. We predict these times can be further improved with a custom hardware that caters more specifically to MCMC processes than a GPGPU.

\begin{figure}[H]
\includegraphics[width=0.5\textwidth]{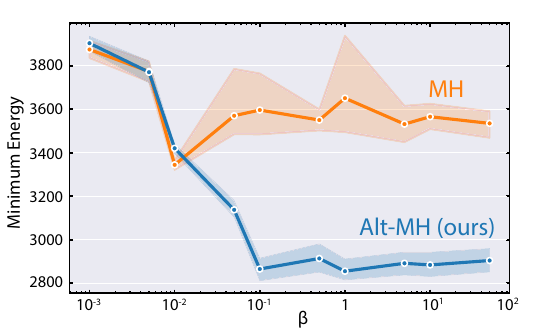}
\caption{\label{fig:beta} Alt-MH is able to find low energy states at exceedingly large $\beta$ for T0955 from CASP 13. $\beta$ is inverse temperature. Energy is measured according to the distogram-only energy function. Each point is the average of the minimum energies of 25 runs, where each run is $82000$ proposals. Shaded bands cover the range between the minimum and maximum values found over 25 runs for each $\beta$. } 
\end{figure}

A crucial benefit of Alternating Metropolis-Hastings is its' diminished dependence on temperature ($1/\beta$) when searching for low energy states. We compare the performance of Alt-MH with traditional MH for varying $\beta$ in Fig.~\ref{fig:beta}. Localized MH prevents Alt-MH from locking into a sub-optimal state at higher $\beta$. This is at odds with the behaviour of traditional Metropolis-Hastings (MH), which can only reach (semi-)low energy states at a specific $\beta$, ex. $\beta \approx 0.01$ in Fig.~\ref{fig:beta}, above or below which the algorithm fails. Alt-MH helps overcome the critical challenge of finding an ideal $\beta$ or annealing schedule for new inputs. 

\section{\label{sec:4}Discussion}

\subsection{\label{sec:4.1} Thoughts on Alternating Metropolis-Hastings}

For Monte Carlo or Markov Chain Monte Carlo methods, visiting states with high internal variation during a random walk generally requires proposal schemes that can generate very specific, low probability states or running for extended periods at a high temperature. An alternative approach is to allow the algorithm to accept states that may increase global energy temporarily but lead to accurate folding of local structures. 

Localized MH fills an ideological role similar to that of noisy restarts in gradient descent-based approaches. When the system converges to some local minimum, it is necessary to introduce noise in such a way that the energy of the system increases temporarily, so that it may once again minimize. Localized MH acts as an intelligent restart, rather than a noisy restart, by adding noise in a way that improves local structures. This makes it easier for future iterations of the minimization scheme to find good global ground states.

A critical benefit of the Alternating-MH algorithm is its diminished dependence on temperature. Fig~\ref{fig:beta} shows that Alt-MH finds low energy states at $\beta$s that far exceed the limits for traditional MH. 

A point to note is that the existing standard for CPU protein simulations in python, pyrosetta, is not built to efficiently find the energy of a subset of a protein structure. Rather the full energy must be calculated, and the subset must be gathered in post. Straightforward changes to the code base could allow for significant improvements in the time required for alternating MH to execute. If Alt-MH were made to run on par with the efficiency that gradient descent based schemes do, then Time To Solution (TTS) measurements could be made on CPU against gradient descent based schemes for the complex, refined energy functions that pyrosetta allows for (trRosetta2).

\subsection{\label{sec:4.2} Other Monte Carlo Approaches to Protein Folding}

Other Monte Carlo approaches to protein folding include population-based schemes like EMC \cite{liang_evolutionary_2001}, genetic algorithms \cite{hansmann_new_1999},  fragment-swapping \cite{park_protein_2018, kandathil_improved_2018,senior_protein_2019}, rotamer substitutions \cite{kuhlman_advances_2019}, simulated annealing \cite{heilmann_sampling_2020}, and multi-particle conformation sampling \cite{wong_exploring_2018}. These methods have largely not been compared to or applied on today’s state of the art protein folding pipelines. Recent works have introduced efficient proposal schemes \cite{cabeza_de_vaca_enhanced_2018} that may integrate well with the Alternating MH approach presented in this work.

Traditional Metropolis-Hastings has been used on all-atom force field equations to sample ground states in a similar manner to this work \cite{heilmann_sampling_2020}. This work also begins at an unfolded protein and creates proposals by twisting dihedral angles of residues in the chain. However, this is not applied to a contact-map based energy function but rather as an alternative to molecular dynamics on a more traditional inter-molecular force field energy function. The idea of localized MH is also absent. Furthermore, this method is heavily temperature dependent, while Alternating MH finds very low energy states without requiring an ideal temperature, or an annealing schedule.  

\subsection{\label{sec:4.3} Future Work}

Incorporation of fragment libraries, novel proposal schemes, annealing schedules, and further testing on parameters like the localization radius $d$ may yield improvements in minimization rate and effectiveness of the Alternating MH algorithm. The Alternating MH algorithm may be applied to other optimization problems like MaxCut or Travelling Salesman Problem (TSP); the key will be to define what a localized segment means in the context of those problems. 

While we show that GPUs are able to accelerate Alt-MH over CPU, we believe even greater improvements in time to solution could be possible with a modified GPGPU that is catered specifically for Monte Carlo and Markov chain Monte Carlo algorithms. Firstly, incorporating some number of local, controllable RNG in each streaming multiprocessor could accelerate the drawing of random numbers. Within the past few years, p-bits built out of stochastic magnetic tunnel junctions have emerged as highly efficient, controllable random number generators that could fill this need~\cite{camsari_stochastic_2017, camsari_p-bits_2019}. We mention the term 'controllable' to motivate interconnected networks of p-bits which may produce random numbers from arbitrary distributions. Secondly, limited shared memory resources hinder block-wise MCMC parallelism. A trade-off of global and local memory for increased shared memory resources would benefit the parallel operation of MCMC algorithms where each block represents an MCMC chain. We believe that these changes would also benefit the implementation of MCMC algorithms for applications extending beyond protein structure prediction. 

\section{\label{sec:5} Conclusions}
We present a novel Markov Chain Monte Carlo method termed Alternating Metropolis-Hastings for optimizing energy functions, applied to protein structure prediction. This algorithm is shown to outperform traditional MH, and traditional MH with noisy restarts handily for 9 proteins taken from recent CASP competitions. We also show that Alt-MH exhibits reduced dependence on temperature compared with traditional MH, a crucial benefit when considering novel problems for which an ideal temperature or schedule is unknown. The algorithm is mapped to a GPGPU to show a 7.5x to 26.5x speedup. Finally, we motivate the development of a modified GPGPU architecture with per-block RNG and increased shared memory resources for accelerated MCMC simulations.

\section*{\label{sec:6} Acknowledgements}
 The authors are grateful for many discussions with Shuvro Chowdhury, Jan Kaiser, and Behtash Behin-Aein over the past year that played a key role in shaping this work. The authors would also like to thank Kerem Camsari and Luke Theogarajan for their advice in the early stages of this work. This work was supported in part by ASCENT, one of six centers in JUMP, a Semiconductor Research Corporation (SRC) program sponsored by DARPA.

\bibliography{refs.bib}
\end{document}